\documentclass{sig-alternate-2013}

\permission{Copyright is held by the International World Wide Web Conference Committee (IW3C2). IW3C2 reserves the right to provide a hyperlink to the author's site if the Material is used in electronic media.}
\conferenceinfo{WWW'14 Companion,}{April 7--11, 2014, Seoul, Korea.} 
\copyrightetc{ACM \the\acmcopyr}
\crdata{978-1-4503-2745-9/14/04. \\
http://dx.doi.org/10.1145/2567948.2579697}

\usepackage{hyperref}
\usepackage{graphicx}
\usepackage{subfigure}

\def\sharedaffiliation{
\end{tabular}
\begin{tabular}{c}}

\begin{document}
\conferenceinfo{COOL}{WWW'14 Workshop Seoul, Korea}

\title{Connecting Dream Networks Across Cultures}

%
%
%
%
%

\numberofauthors{2} 
%

\author{
%
%
\alignauthor Onur Varol\titlenote{Corresponding author: \texttt{ovarol@indiana.edu}}\\
\alignauthor Filippo Menczer\\
\sharedaffiliation
\affaddr{Center for Complex Networks and Systems Research}\\
\affaddr{School of Informatics and Computing, Indiana University, Bloomington, USA}
}


\maketitle

\begin{abstract}
Many species dream, yet there remain many open research questions in the study of dreams. 
The symbolism of dreams and their interpretation is present in cultures throughout history.
Analysis of online data sources for dream interpretation using 
network science leads to understanding symbolism in dreams and their associated meaning. 
In this study, we introduce dream interpretation networks for English, Chinese and 
Arabic that represent different cultures from various parts of the world. We 
analyze communities in these networks, finding that symbols within a community are 
semantically related. The central nodes in communities give insight about cultures and symbols 
in dreams. The community structure of different networks highlights cultural similarities and 
differences. Interconnections between different networks are also identified by translating symbols
from different languages into English. Structural correlations across networks point out 
relationships between cultures. Similarities between network communities are also investigated 
by analysis of sentiment in symbol interpretations. We find that interpretations within a 
community tend to have similar sentiment. Furthermore, we cluster communities based on 
their sentiment, yielding three main categories of positive, negative, and neutral dream 
symbols.
\end{abstract}





\newpage
\section{Introduction}

Systems in society and nature contain connections that capture
useful information and provide insights for our understanding. In particular, representing
knowledge as a network offers depth to our understanding of nature and society. 
Components of these systems can be mapped as a network to analyze. For example 
networks of drugs~\cite{yildirim2007drug} and diseases~\cite{goh2007human} 
help us to design new drugs and learn more about human diseases. Networks of products
are also studied for modeling markets and the economy~\cite{hidalgo2007product}. 
Using networks to model brain connectivity helps to understand functional parts of brains and 
mechanisms of these functional regions~\cite{bullmore2012economy}. Networks  are also useful for understanding
human mobility, social relations, political discussion~\cite{Truthy_icwsm2011politics}, human dynamics, and society on a
large scale~\cite{brockmann2006scaling,szell2010multirelational,ferrara2013traveling}. 
As network science evolves, it also can be  seen in novel areas such as cuisine, 
where it is being used to characterize the eating habits of different cultures~\cite{ahn2011flavor}, 
and science of science, where it contributes to our understanding of the emergent 
relationships between disciplines, scholars, and papers~\cite{Sun2013srep}.

\emph{Interconnected networks} (with multiple connected layers) recently have attracted the interest of the  scientific community. 
Researchers have investigated statistical mechanics~\cite{bianconi2013statistical} and 
phase transitions of multiplex networks~\cite{radicchi2013abrupt} (where  there is a 1-1 correspondence between nodes in different layers), 
diffusion dynamics~\cite{gomez2013diffusion,saumell2012epidemic}, and 
emergence of these networks~\cite{nicosia2013growing}. Various networks in society 
can be represented as multiplex networks. Our goal is to advance \emph{oneirology}  
(the scientific study of dreams) by building an interconnected network of dream interpretations. 

In spite of the universal and common experience of dreaming, its purpose and mechanisms are still 
largely unknown. One of the earliest scientific studies in 
oneirology dates from about 300 years ago~\cite{dreams}. 
Throughout  history people have recorded dreams in an attempt
to explain their connection with waking life~\cite{hall1953cognitive}. 
Content analysis of dreams was studied by psychologists Sigmund Freud, Carl Jung, and 
Calvin Hall~\cite{freud2004interpretation,jung1963memories,hall1953cognitive}. 
While Freud thought dreams to be aimed at  unconscious fulfillment and explained symbols
through metaphor, Jung argued that the
purpose of dreams were unconscious messages to the self. Calvin Hall  
collected more than 50,000 dream reports and annotated more than 1,000 of them. 
He found that dreams all over the world contain similar 
concepts: the self, other people, and situations~\cite{hall1966content}. Recently
codebooks for dream interpretation and collections of dreams have become available online 
in the DreamBank database.\footnote{\url{http://DreamBank.net}} 
Dream contents have been analyzed to investigate similarities
between dreams and waking life, observing common terms about religion and 
sexuality~\cite{domhoff2008studying}. Schweickert~\cite{schweickert2007properties} 
built and compared networks of individuals in dreams and waking life.

In this study, we collect online data from dream dictionaries and analyze it 
using network analysis techniques to unveil communities in dream interpretation
networks and associations between symbols.

\subsection{Contributions and outline}

In the remainder of the paper we make the following contributions:

\begin{itemize}
	\item We discuss how we collected and built dream interpretation 
	datasets for three different languages as introduced in \S~\ref{sub:dataset} 
	and create the first dream interpretation network as described in \S~\ref{sub:build_network}. 
	
	\item In \S~\ref{sub:multicultural} we show how to build an interconnected multicultural 
	dream network by using dictionary data for dream symbols. 	
	
	\item \S~\ref{sub:communities} reports on community detection in dream interpretation
	networks, and on our analysis of the characteristics of the resulting communities. The central nodes in these communities are inspected to grasp their topics, finding that representative nodes in the same community convey similar messages.
	
	\item In \S~\ref{sub:layer_correlation} we measure node strength and edge weight correlations  between 
	layers, finding that dream symbols in different languages have positively correlated properties.
	
	\item Finally, \S~\ref{sub:sentiment_analysis} highlights the role of sentiment in similar communities in the 
	English network and the clustering of communities with similar happiness scores.
	
\end{itemize}

\section{Experimental setup}
In this section we describe the methodology we followed
to collect three datasets of dream interpretations in different
cultures and to construct content similarity networks that
allow us to study cultural aspects of dream interpretation. 

\subsection{Datasets}
\label{sub:dataset}

To build our dream interpretation datasets for different cultures,
we crawled websites of online dream dictionaries from various sources. 
The crawled Web pages were parsed to extract dreams symbols 
with their associated interpretations.

Pre-processing techniques specific to each language (described next) 
were applied to the raw data extracted from Web pages in that language.   
Information retrieval techniques were then used to convert text into vector 
space representations, used for building dream interaction networks. 
To study dreams across different cultures, we selected the three languages 
and sources below.

\paragraph{English}
The dream dictionary that we crawled\footnote{\url{http://www.dreamdictionary.org}} contains 
1,391 distinct symbols (English terms). We populated our dataset with these 
symbols and their interpretations. Interpretations of symbols were pre-processed using lemmatization, 
Porter stemming~\cite{porter1980algorithm}, and removal of stop-words.

\paragraph{Chinese}
We crawled the content of a 
website\footnote{\url{http://zgjm.xixik.com}} for traditional Chinese
dream interpretations and collected the interpretations of about
1,140 distinct symbols.
Unlike many languages, Chinese is written without using spaces to separate words. 
In Chinese text retrieval, segmentation is an important pre-processing step. 
We used software for segmentation of Chinese
text\footnote{\url{https://github.com/fxsjy/jieba}} and removed unicode punctuation 
characters in Chinese.

\paragraph{Arabic}
We collected dream symbols and interpretations
in Arabic from a dictionary website\footnote{\url{http://dreams.svalu.com}} 
containing 2,419 distinct symbols.
Arabic retrieval tasks require some effort to clean text data. We used the ISRI 
Stemmer~\cite{taghva2005arabic} integrated in the NLTK~\cite{bird2009natural} 
software package, which enables the removal of stop-words, language-specific 
dialectics, initial characters, and prefixes.

\subsection{Building networks}
\label{sub:build_network}

To investigate relations between dream symbols and make comparisons
between different cultures, we built dream interpretation networks in each 
language. These networks are weighted and undirected. Each node $i$ corresponds
to a symbol in a dream dictionary and each weight $e_{ij}$ to the similarity between 
interpretation documents for symbols $i$ and $j$, respectively. 
Higher weights represent similar interpretations and closer meanings.

For the computation of document similarities we employed a commonly used vector space 
representation, namely the TF-IDF~\cite{Jones1973619} vector $d_i$ 
for the interpretation document of each symbol $i$. 
Similarities between document representations are computed by the cosine similarity
\begin{equation}
e_{ij} = \frac{\sum_{w \in W_i \cap W_j} d_{iw} d_{jw}}{\sqrt{\sum_{w \in W_i} d_{iw}^2}\sqrt{\sum_{w \in W_j} d_{jw}^2}},
\end{equation}
where $d_{iw}$ is the TF-IDF weight assigned to term $w$ in document vector $d_i$ and $W_i$ is the set of words in $d_i$. 

To analyze the resulting weighted networks, we applied a multi-scale backbone
extraction technique~\cite{serrano2009extracting} to remove statistically 
insignificant edges. In this algorithm, the significance level of edges is controlled
by a parameter $\alpha$. We tuned $\alpha$ for each network to obtain a minimal backbone, 
i.e, a network containing the minimum number of edges such that all nodes are in a single 
connected component. The structural properties of the resulting networks 
are summarized in Table~\ref{table:network_statistics}. 
The networks are of course very different, however they display some structural similarities. All of them have high density and clustering 
coefficients, and short path lengths. In other words, they are small-world networks~\cite{watts1998collective}.

\begin{table}[!h]
\caption{Structural properties of dream interpretation networks generated by
backbone extraction for appropriate values of the parameter $\alpha$.}
\centering
	\begin{tabular}{p{3cm}ccc} 
	\hline	
	Network & English & Chinese & Arabic \\
	&($\alpha=.05$)&($\alpha=.15$)&($\alpha=.05$) \\
	\hline
	\# nodes & 1391 & 1140 & 2419 \\
	\# edges & 46302 & 39864 & 55893 \\
	Density & 0.048 & 0.061 & 0.012 \\	
	Avg. degree & 66.57 & 69.94 & 46.21 \\
	Avg. strength & 6.26 & 6.91 & 6.68 \\
	Clustering coefficient & 0.15 & 0.21 & 0.17 \\
	Avg. shortest path & 2.05 & 2.05 & 2.51 \\
	\hline
	\end{tabular}
\label{table:network_statistics}
\end{table}	

\begin{figure*}[t!]
	\centerline{
		\includegraphics[width=0.33\textwidth]{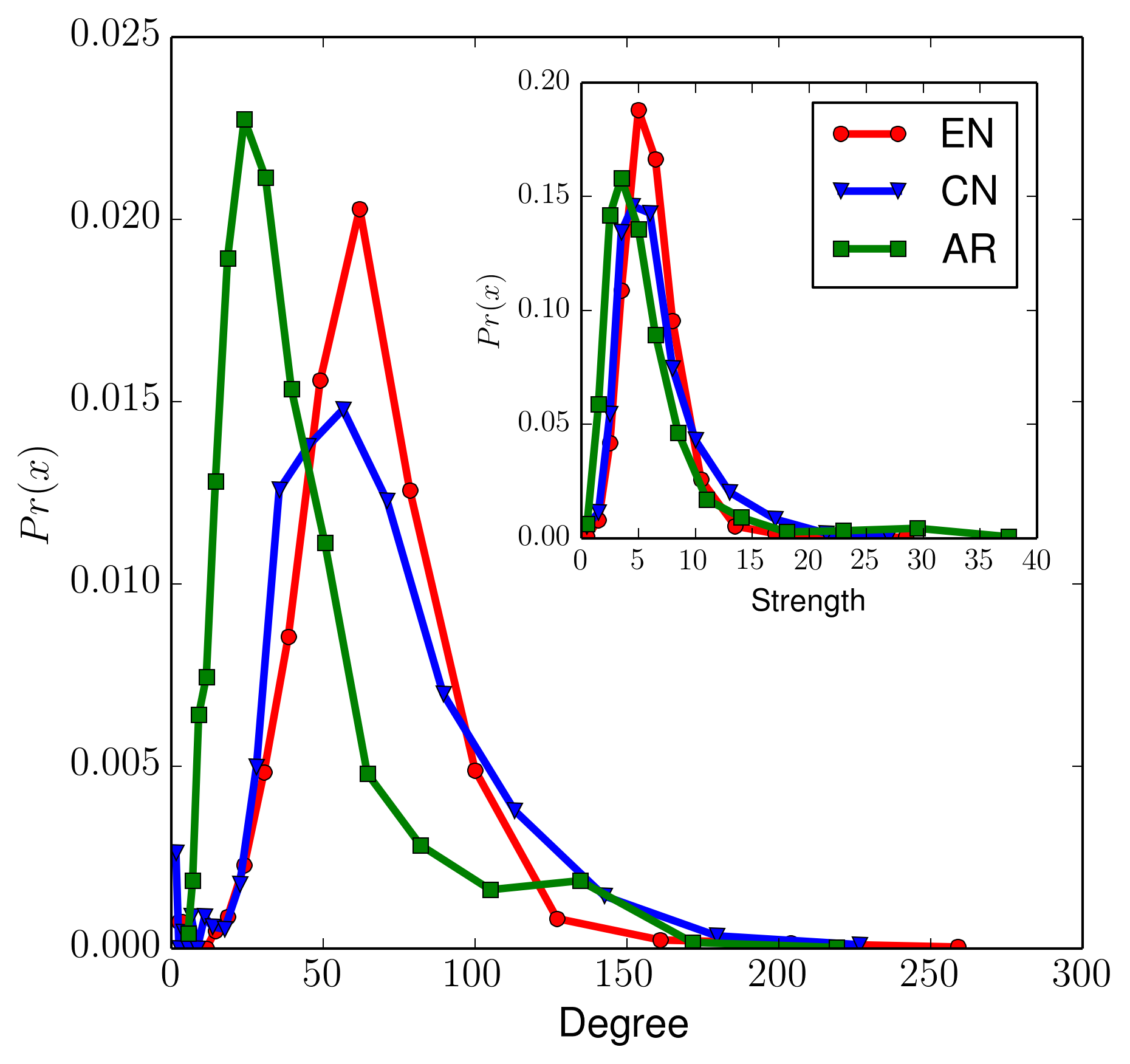}
		\includegraphics[width=0.33\textwidth]{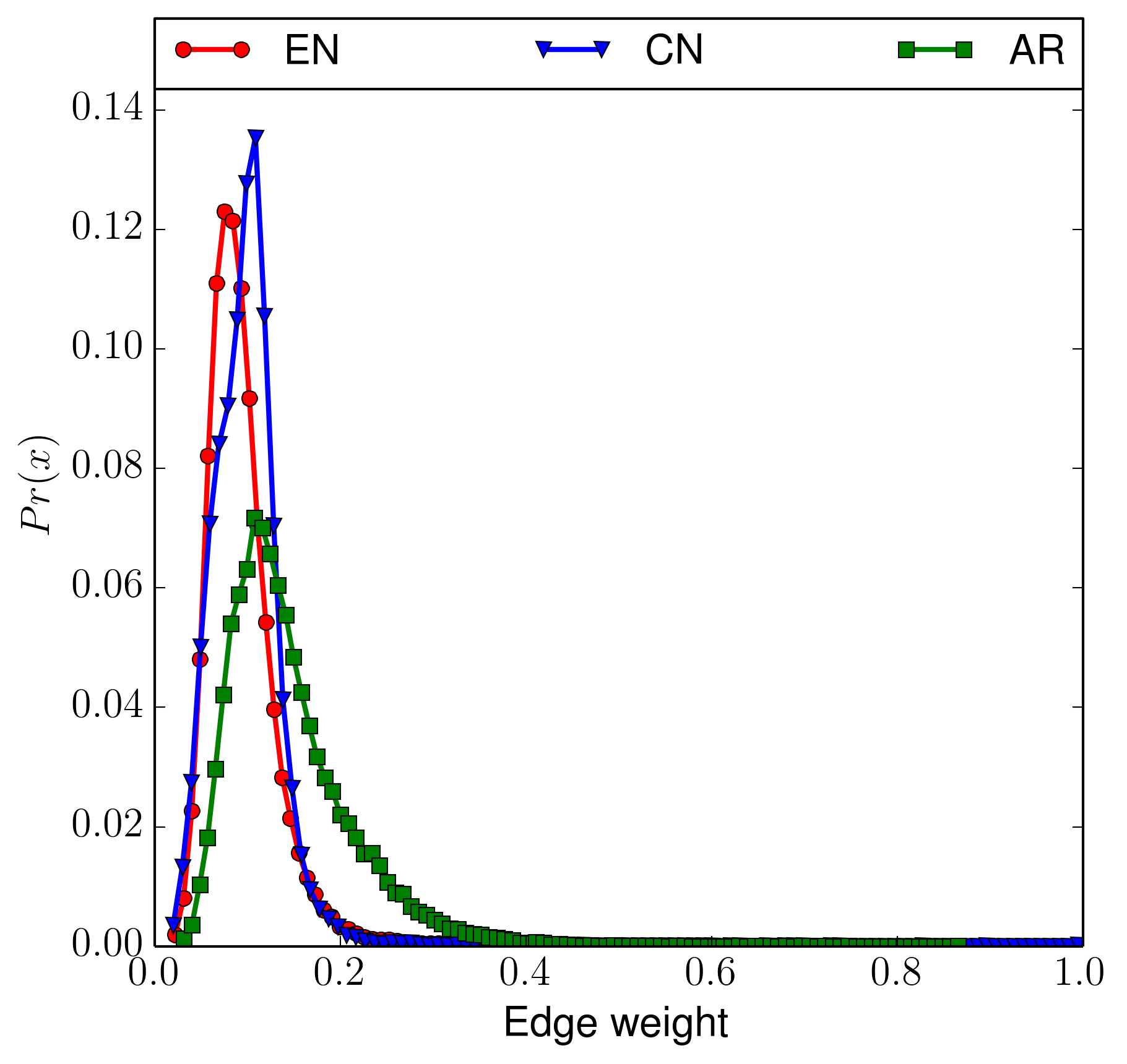}
		\includegraphics[width=0.33\textwidth]{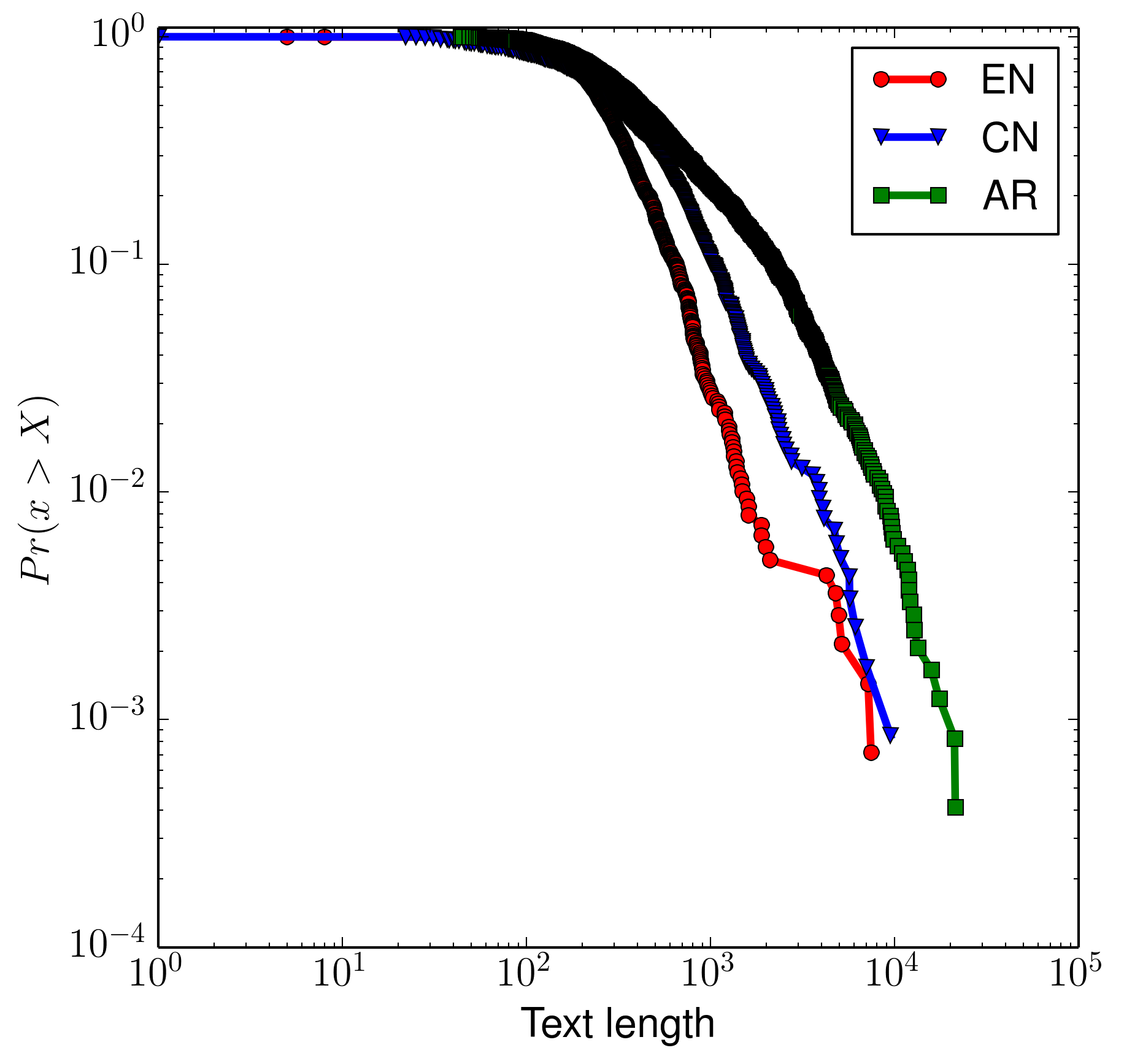}
	}
	\caption{Distributions of network properties: (a) degree and strength (inset), (b) edge weight, and (c) textual length of dream symbol interpretations (CDF).}
	\label{fig:content_distribution}
\end{figure*}

Fig.~\ref{fig:content_distribution} displays distributions of a few network properties. 
Similar characteristics are observed in the networks corresponding to different languages. 
The distributions of degree, strength (weighted degree), and edge weights are narrow (Poissonian), while text length has a skewed distribution spanning several orders of magnitude. 

\newpage
\subsection{Multicultural dream network}
\label{sub:multicultural}

Building interconnected networks requires a set of nodes appearing in multiple dream interpretation
networks. Direct translation is the easiest method for matching nodes in different networks.
We used Google's translation service to translate Arabic and Chinese symbols into English, and employed the English terms to match symbols in different layers.

\section{Results}

In this section we carry out community detection for the dream interpretation networks and 
compute correlations between multiple layers. 
We observe correlations that points to similarities between
different cultures. Finally, we explore the role of emotions in dream interpretations 
by performing sentiment analysis in the English network.

\subsection{Communities}
\label{sub:communities}

Communities provide a more compact representation of large networks~\cite{fortunato2010community}. 
The identification of the central nodes in a community allows us
to recognize the role or meaning of the community.

We detected communities in dream interpretation networks by 
using the Louvain algorithm~\cite{blondel2008fast}, 
as implemented by Guillaume~\cite{louvain}, which performs 
a greedy optimization of Newman's modularity~\cite{newman2004finding} on a weighted network. The Louvain method was executed
500 times for each network to find the best modularities: 0.24, 0.37, and 0.44 
for English, Chinese, and Arabic, respectively. 

\begin{figure*}
\centerline{
	\includegraphics[width=0.33\textwidth]{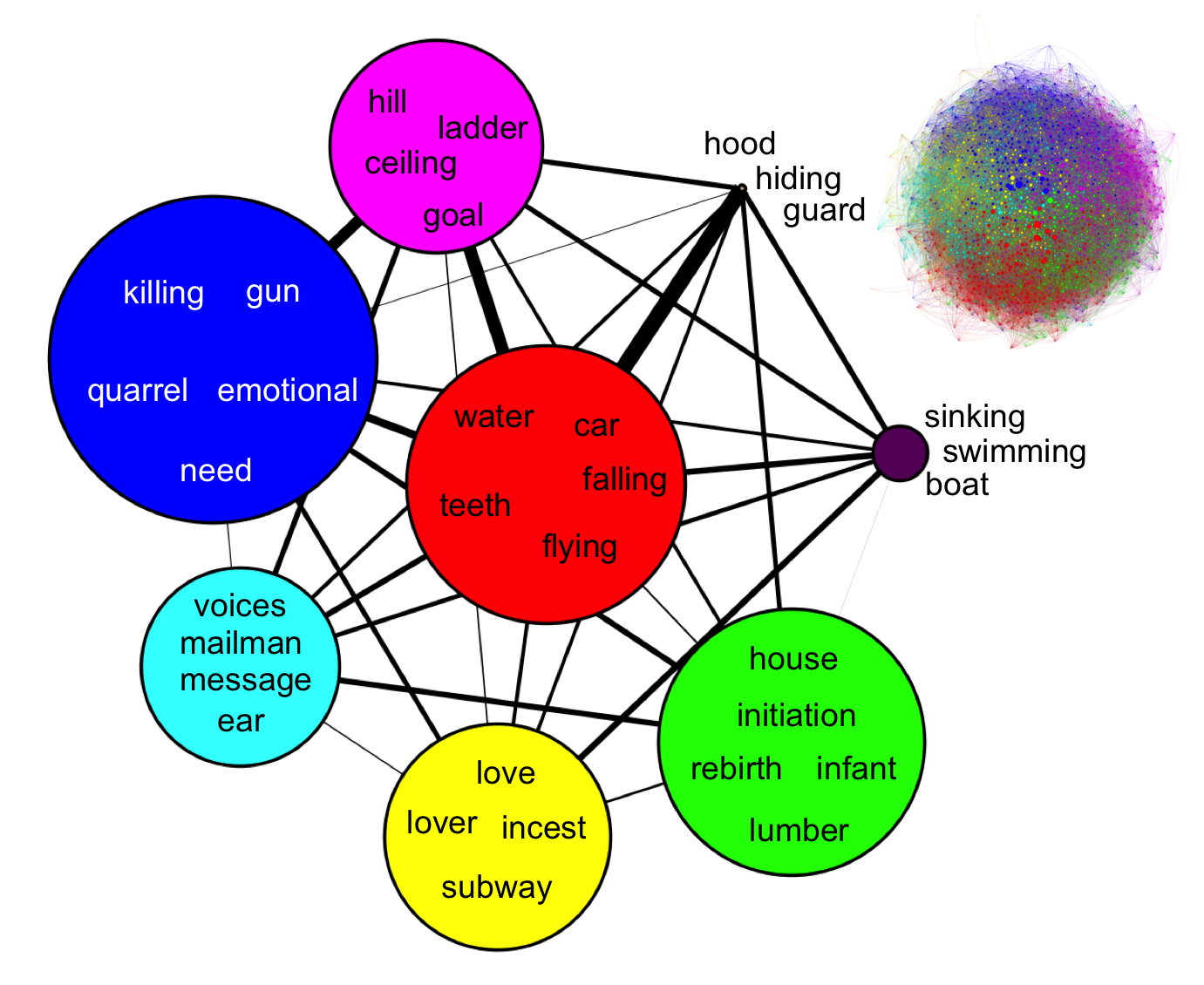}
	\includegraphics[width=0.33\textwidth]{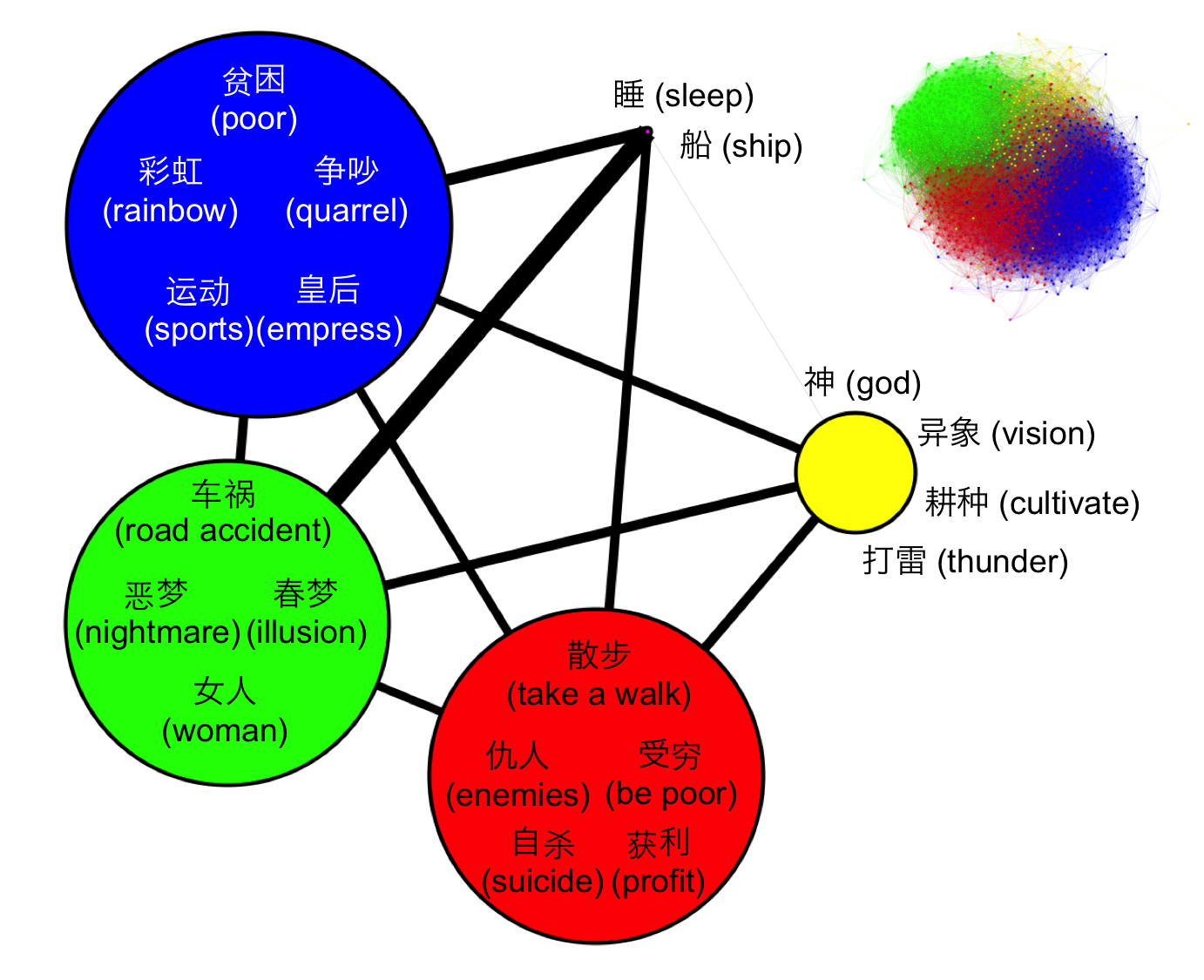}
	\includegraphics[width=0.33\textwidth]{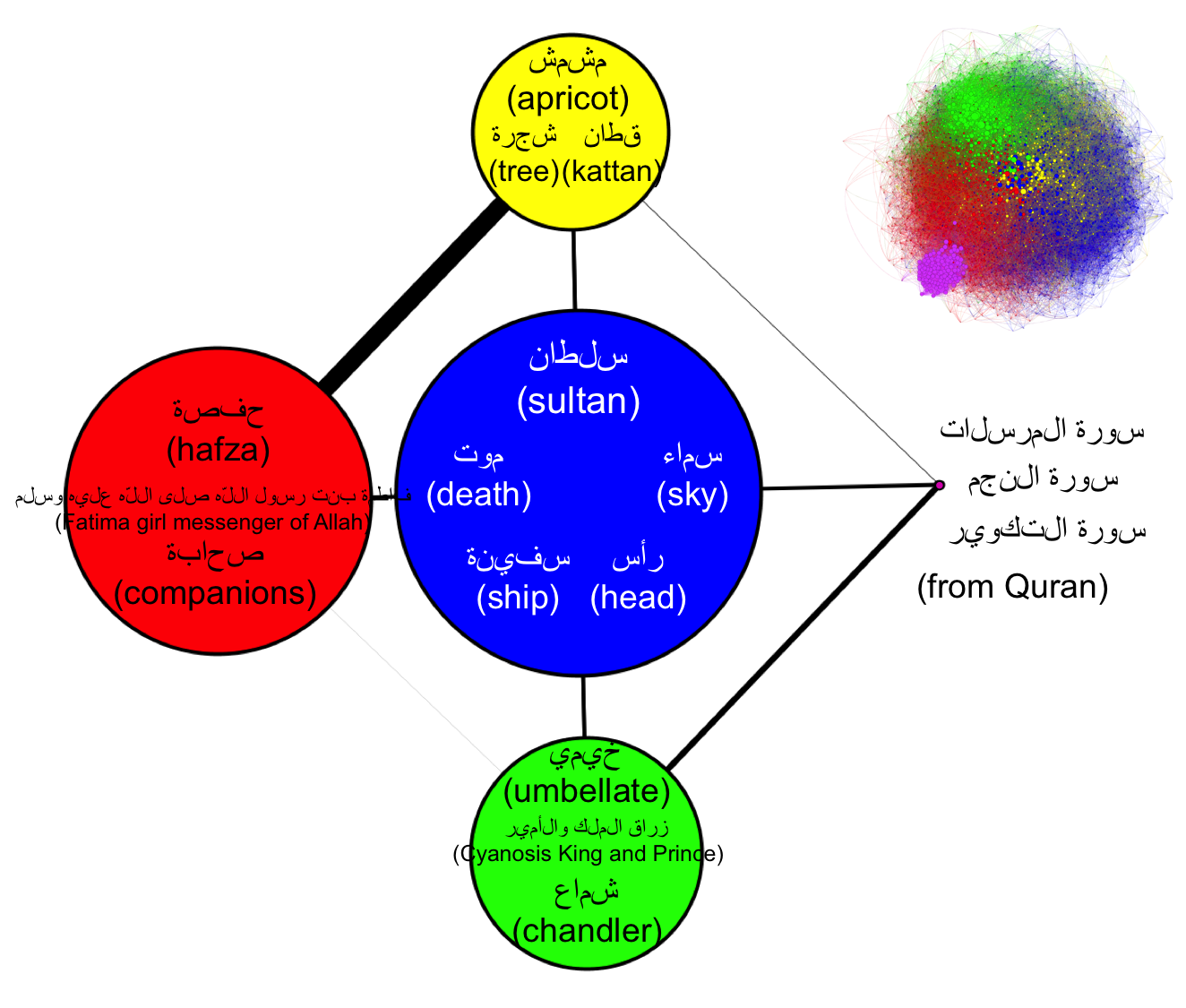}
}
	\caption{Communities detected by Louvain method: (a) English, (b) Chinese, and (c) Arabic. To label each community, we list the symbols with the highest Eigenvector centrality. The complete networks are illustrated in the upper right corners of each figure, with colors corresponding to the communities (colors are not intended to match across different networks).}
	\label{fig:communities}
\end{figure*}

Since the networks are built from symbol similarities, it is natural to expect 
that a community should join related symbols corresponding to a shared topic. 
To identify topics, we extract the central nodes in each community using the 
eigenvector centrality algorithm, which gives credit to a node by considering both 
its degree and that of its neighbors~\cite{bonacich2007some}. Fig.~\ref{fig:communities}
depicts coarse-grained networks in which nodes represent the communities. An edge between two community nodes has a weight equal to the total weight of the links connecting symbols 
in the two communities. 

According to Hall's cognitive theory of dream symbols~\cite{hall1953cognitive},  
we expected to observe communities containing dream symbols that serve either 
as warning for the self or reactions to daily life and others. Automatic translation 
of the symbols helps to determine if these meanings are preserved across networks. 

In the English network (Fig.~\ref{fig:communities}(a)), the community that contains `goal,' `hill,' and `ladder' is interpreted as achievement after a struggle, while the community labeled `voices,' `mailman,' and `message' represents warning and precaution. The community with symbols `hiding,' `guard,' and  `raincoat' means having protection. 

The Chinese community in Fig.~\ref{fig:communities}(b) that contains the symbols `rainbow,' 
`poor,' and `quarrel' tells about good things that will happen in the future, while the one with the nodes `road accident,' `woman,' and `illusion' contains symbols about forewarning and need of life changes. The community labeled `suicide,' `walking,' and `enemies' 
is interpreted as describing happy life and end of troubles. 
Note the rough correspondence between the first two example communities mentioned for English and Chinese networks.

In the Arabic network represented in Fig.~\ref{fig:communities}(c), the role of religious characters and objects are identifiable in distinct communities represented in purple and red. Similarly in the Chinese network, the yellow community contains symbols like `god' and `vision.' Content analysis of symbols in DreamBank leads us to similar conclusions about religious dreams~\cite{domhoff2008studying}.

Given the low reliability of automatic translation, we do not want to make any strong claims about correspondence between interpretations in different cultures. However, we can compare the networks by investigating correlation between strengths and edge weights.

\subsection{Correlation between layers}
\label{sub:layer_correlation}

One way to look at the structural similarities between dream interpretation networks in different languages is to focus on common nodes (detected by automatic symbol translation) and measure the correlations between properties of these nodes. Let's consider two properties: strength of common nodes and weight of edges connecting pairs of common nodes. We measure Pearson's correlation $r$. For  strength the definition is straightforward. For edge weight the correlation is defined as follows:
\[
r(A,B) = \frac{\displaystyle\sum_{i,j \in S^A\cap S^B} ( e_{ij}^A - \overline{e^A} ) ( e_{ij}^B - \overline{e^B} ) }{ \sqrt{\displaystyle\sum_{i,j \in S^A} ( e^A_{ij} - \overline{e^A}  )^2} \sqrt{\displaystyle\sum_{i,j \in S^B} ( e^B_{ij} - \overline{e^B}  )^2}}, 
\]
where $\overline{e^A}$ is the average weight of edges in network $A$. We only consider edges between  nodes common to both networks. $S^A$ is the set of edges among common nodes that are present in $A$.

Table~\ref{table:interconnections} reports on the numbers of common nodes and edges, as well as correlations between layers. The correlations are not very strong, but positive and significant in almost all cases. These results indicate that interconnected nodes have somewhat similar connectivity: symbols tend to be connected to the same neighbors across languages. Symbols also tend to have similar strength: general (specific) symbols in one language tend to be general (specific) in other languages as well. These results hold whether we consider the backbone or fully connected networks.

\begin{table*}[t!]
\caption{Pearson correlation for pairs of common nodes (interconnections) identified by translation of symbols. We compute the correlation of edge weight and node strength over two types of network: fully connected and backbone. The correlation values that are statistically significant ($p<0.05$) are marked in bold, based on the sample sizes (common nodes or edges).}
\label{table:interconnections}
\centering
	\begin{tabular}{p{3cm} c c c c c c c}
	\hline
	Networks & Common nodes & \multicolumn{2}{ c }{Common edges} & \multicolumn{2}{ c }{Node strength $r$} & \multicolumn{2}{ c }{Edge weight $r$}\\	
	\cline{3-8}
	& & Backbone & Full & Backbone & Full & Backbone & Full\\
	\hline
	Arabic \& Chinese & 	 378 & 107 & 68,162 & 	\textbf{0.09} & 0.06 &	\textbf{0.24} & \textbf{0.04} 	\\ 	
	Chinese \& English & 274 & 151 & 32,196 & 	\textbf{0.20} & \textbf{0.13} &	\textbf{0.15} & \textbf{0.05} 	\\ 	 
	Arabic \& English & 	 239 & 39 & 26,193 	& 	0.10 & \textbf{0.23} &	\textbf{0.43} & \textbf{0.12} \\
	\hline
	\end{tabular}
\end{table*}

\subsection{Sentiment analysis}
\label{sub:sentiment_analysis}

Sentiment analysis is an important tool for probing textual data to identify
nuanced differences, such as happiness. In dreams, emotions
and feelings play important roles in the interpretation of the meaning of symbols. 
We analyzed the symbol interpretations in each community to study whether communities 
can be further grouped into larger clusters based on similar sentiment. We computed a happiness
score for each symbol using a dataset of happiness scores in 
English~\cite{kloumann2012positivity}. This dataset contains happiness scores for
more than 10,000 English words, each ranging between 1 and 9 (1: least happy, 5: neutral, 
and 9: most happy). Each interpretation 
text contains a set of words that has a happiness score in the dataset. The happiness
score for the symbol is averaged across the words in the interpretation text.

\begin{figure}[b!]
	\centering
	\includegraphics[width=\columnwidth]{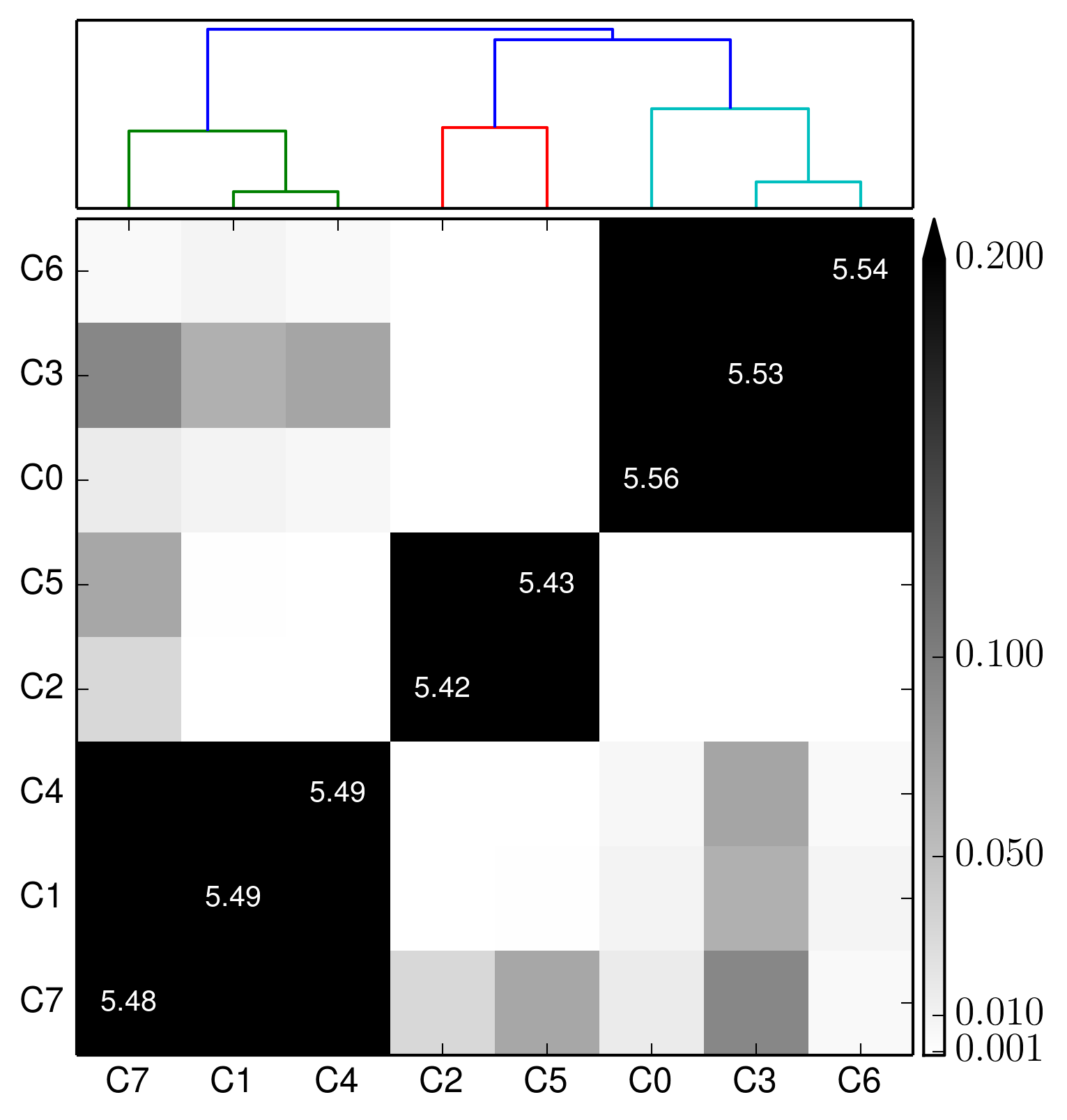}
    \caption{Hierarchical clustering of communities in the English dream interpretation network, using symbol happiness distributions to compute similarities (darker means more similar). The values on the diagonal 
    represent average happiness scores within each community.}
    \label{fig:sentiment}
\end{figure}

We further investigated the sentiment properties of the dream network. Building a sentiment similarity network among symbols is infeasible, because we cannot evaluate statistical significance of the similarity between two individual sentiment scores. Therefore, we focused on the relationships between
sentimentally similar communities. The symbols in each community produce happiness score samples 
drawn from the distribution of sentiment. These samples allow us to compute 
statistical significance ($p$-values) of differences between community distributions, using 
two-sample $t$-tests. 
The $p$-values are then used as similarity measures between communities, and passed 
to an aggregative hierarchical clustering algorithm to join sentimentally similar communities.  
Fig.~\ref{fig:sentiment} reports the similarities between communities and illustrates the dream clusters with similar sentiment.  
A multiple comparison Kruskal-Wallis test~\cite{kruskal1952use} of sentiment scores confirms that the clusters are highly significant.
The results are not particularly dependent on the similarity measure; Jensen-Shannon divergence~\cite{lin1991divergence} between the distributions yields similar partitions.  

The most central symbols in the sentimentally similar communities are illustrated in   
Table~\ref{table:community_central_nodes}. 
Each community in the English dream interpretation network deserves investigation. 
Interpretations of symbols in C1 concern fresh starts and new beginnings; 
C4 mostly deals with warnings about upcoming 
 life changes and suggests awareness to surrounding events. C7 contains symbols that describe forms of protection. These three communities concern change, forewarning, and protection from the unexpected. Similarly, we analyzed the symbols in 
communities C2 and C5, which have the lowest sentiment scores. These symbols deal with  troubles in life and 
personal relationships and encountering highly stressful situations. The last group of communities 
has the highest sentiment scores. 
Symbols in communities C0 and C3 are associated with strong emotions about specific situations and others.  C6 contains symbols dealing with  ambition for reaching targets or achieving success. 

In all of these clusters of communities, sentiment plays a meaningful and binding role.
Inspection of the central nodes in the Arabic and Chinese network communities 
suggests roughly similar relations, in agreement with Hall's
cognitive theory on dream symbols~\cite{hall1953cognitive}. 

\begin{table}
\caption{Communities in the English dream interpretation network, grouped by
their sentiment similarities. The groups are identified by hierarchical clustering.
The symbols  with the highest eigenvector centrality are used to label each community.}
\centering
	\begin{tabular}{c | p{6cm}} 
	\hline
	Community  & Central symbols \\
	\hline
	\hline
	C1 & house, rebirth, initiation, infant, lumber\\ 
	C4 & voices, mailman, message, fax machine, ear\\ 
	C7 & hiding, overcoat, guard, hood, raincoat\\ 
	\hline
	C2 & falling, car, teeth, flying, water\\ 
	C5 & gun, need, killing, quarrel, emotional\\ 
	\hline
	C0 & houseboat, boat, sinking, swimming, scuba diving\\ 
	C3 & love, lover, subway, love note, incest\\ 
	C6 & goal, ceiling, hill, ladder, rock climbing\\ %
	\hline
	\end{tabular}
\label{table:community_central_nodes}
\end{table}

\section{Conclusions}

In this work we introduced the concept of a dream interpretation network.
Our study bridges network science and oneirology; 
properties of communities were investigated using network science and 
sentiment analysis techniques. 
Our analysis was carried out on multiple networks in different languages 
to investigate the role of culture in the associations of dream symbols. 
We sought to understand communities in dream interpretation networks, and our 
findings support Hall's cognitive theory of dreams.  

We built a multicultural dream network by identifying interconnections between language layers.
Nodes with interconnections are sparse compared to the overall network due to our 
reliance on automatic translation to match symbols in different languages. We also 
explored the identification of interconnections by using a Thesaurus Web service 
available online.\footnote{\url{http://thesaurus.altervista.org}} Such an approach 
did not improve the correlations between layers, possibly because this Web service 
does not sort synonyms by their confidence.  In the future, it would be interesting to investigate
cross-cultural relations by using other online resources, such as Wikipedia.  

A more detailed analysis with different algorithms of community detection in the multilayer symbols network is left for future studies. Symbols such as 
`water,' `teeth,' and `flying' have multiple explanations depending
on actors or associated colors and other context. They will require extensive analysis for 
deeper understanding. Content analysis of user-reported dreams have been carried 
out to identify words associated with diverse emotions~\cite{domhoff2008studying}. 
It would also be desirable to study the co-occurence of symbols in dreams to derive 
a better understanding of associations between symbols. 

\section{Acknowledgments}
We are grateful Yong-Yeol Ahn, the NaN group 
(\url{cnets.indiana.edu/groups/nan}), and three anonymous reviewers for helpful feedback, 
and to John McCurley for editing assistance. 
This work is partially supported by DARPA
(grant W911NF-12-1-0037), NSF (grant CCF-1101743), and the McDonnell Foundation.
The funders had no role in study design, data collection and analysis, decision to publish, or preparation of the manuscript.

\newpage
\bibliographystyle{plain}
\bibliography{sigproc}

\begin{thebibliography}{10}

\bibitem{ahn2011flavor}
Yong-Yeol Ahn, Sebastian~E Ahnert, James~P Bagrow, and Albert-L{\'a}szl{\'o}
  Barab{\'a}si.
\newblock Flavor network and the principles of food pairing.
\newblock {\em Scientific Reports}, 1, 2011.

\bibitem{bianconi2013statistical}
Ginestra Bianconi.
\newblock Statistical mechanics of multiplex networks: Entropy and overlap.
\newblock {\em Physical Review E}, 87(6):062806, 2013.

\bibitem{bird2009natural}
Steven Bird, Ewan Klein, and Edward Loper.
\newblock {\em Natural language processing with Python}.
\newblock O'Reilly, 2009.

\bibitem{blondel2008fast}
Vincent~D Blondel, Jean-Loup Guillaume, Renaud Lambiotte, and Etienne Lefebvre.
\newblock Fast unfolding of communities in large networks.
\newblock {\em Journal of Statistical Mechanics: Theory and Experiment},
  2008(10):P10008, 2008.

\bibitem{bonacich2007some}
Phillip Bonacich.
\newblock Some unique properties of eigenvector centrality.
\newblock {\em Social Networks}, 29(4):555--564, 2007.

\bibitem{brockmann2006scaling}
Dirk Brockmann, Lars Hufnagel, and Theo Geisel.
\newblock The scaling laws of human travel.
\newblock {\em Nature}, 439(7075):462--465, 2006.

\bibitem{bullmore2012economy}
Ed~Bullmore and Olaf Sporns.
\newblock The economy of brain network organization.
\newblock {\em Nature Reviews Neuroscience}, 13(5):336--349, 2012.

\bibitem{Truthy_icwsm2011politics}
Michael Conover, Jacob Ratkiewicz, Matthew Francisco, Bruno Gon\c{c}alves,
  Alessandro Flammini, and Filippo Menczer.
\newblock Political polarization on twitter.
\newblock In {\em Proc. 5th International AAAI Conference on Weblogs and Social
  Media (ICWSM)}, 2011.

\bibitem{domhoff2008studying}
G~William Domhoff and Adam Schneider.
\newblock Studying dream content using the archive and search engine on
  dreambank. net.
\newblock {\em Consciousness and Cognition}, 17(4):1238--1247, 2008.

\bibitem{ferrara2013traveling}
Emilio Ferrara, Onur Varol, Filippo Menczer, and Alessandro Flammini.
\newblock Traveling trends: social butterflies or frequent fliers?
\newblock In {\em Proc. of 1st ACM Conference on Online Social Networks
  (COSN)}, pages 213--222. ACM, 2013.

\bibitem{fortunato2010community}
Santo Fortunato.
\newblock Community detection in graphs.
\newblock {\em Physics Reports}, 486(3):75--174, 2010.

\bibitem{freud2004interpretation}
Sigmund Freud.
\newblock {\em The interpretation of dreams}.
\newblock Kessinger Publishing, 2004.

\bibitem{goh2007human}
Kwang-Il Goh, Michael~E Cusick, David Valle, Barton Childs, Marc Vidal, and
  Albert-Laszlo Barabasi.
\newblock The human disease network.
\newblock {\em Proceedings of the National Academy of Sciences},
  104(21):8685--8690, 2007.

\bibitem{gomez2013diffusion}
S.~Gomez, A.~Diaz-Guilera, J.~Gomez-Gardenes, C.~J. Perez-Vicente, Y.~Moreno,
  and A.~Arenas.
\newblock Diffusion dynamics on multiplex networks.
\newblock {\em Phys. Rev. Lett.}, 110:028701, Jan 2013.

\bibitem{hall1953cognitive}
Calvin~S Hall.
\newblock A cognitive theory of dream symbols.
\newblock {\em The Journal of General Psychology}, 48(2):169--186, 1953.

\bibitem{hall1966content}
Calvin~S Hall and Robert~L Van~de Castle.
\newblock {\em The content analysis of dreams}.
\newblock Appleton-Century-Crofts, 1966.

\bibitem{hidalgo2007product}
C{\'e}sar~A Hidalgo, Bailey Klinger, A-L Barab{\'a}si, and Ricardo Hausmann.
\newblock The product space conditions the development of nations.
\newblock {\em Science}, 317(5837):482--487, 2007.

\bibitem{Jones1973619}
Karen~Sparck Jones.
\newblock Index term weighting.
\newblock {\em Information Storage and Retrieval}, 9(11):619--633, 1973.

\bibitem{jung1963memories}
Carl~G Jung.
\newblock Memories, dreams.
\newblock {\em Reflections}, 84, 1963.

\bibitem{kloumann2012positivity}
Isabel~M Kloumann, Christopher~M Danforth, Kameron~Decker Harris, Catherine~A
  Bliss, and Peter~Sheridan Dodds.
\newblock Positivity of the english language.
\newblock {\em PloS One}, 7(1):e29484, 2012.

\bibitem{kruskal1952use}
William~H Kruskal and W~Allen Wallis.
\newblock Use of ranks in one-criterion variance analysis.
\newblock {\em Journal of the American Statistical Association},
  47(260):583--621, 1952.

\bibitem{louvain}
Etienne Lefebvre and Jean-Loup Guillaume.
\newblock Louvain method: Finding communities in large networks.
\newblock \url{sites.google.com/sites/findcommunities}, 2008.

\bibitem{lin1991divergence}
Jianhua Lin.
\newblock Divergence measures based on the shannon entropy.
\newblock {\em IEEE Transactions on Information Theory}, 37(1):145--151, 1991.

\bibitem{newman2004finding}
Mark~EJ Newman and Michelle Girvan.
\newblock Finding and evaluating community structure in networks.
\newblock {\em Physical Review E}, 69(2):026113, 2004.

\bibitem{nicosia2013growing}
V.~Nicosia, G.~Bianconi, V.~Latora, and M.~Barthelemy.
\newblock Growing multiplex networks.
\newblock {\em Phys. Rev. Lett.}, 111:058701, Jul 2013.

\bibitem{porter1980algorithm}
Martin~F Porter.
\newblock An algorithm for suffix stripping.
\newblock {\em Program: Electronic library and information systems},
  14(3):130--137, 1980.

\bibitem{radicchi2013abrupt}
Filippo Radicchi and Alex Arenas.
\newblock Abrupt transition in the structural formation of interconnected
  networks.
\newblock {\em Nature Physics}, 2013.

\bibitem{dreams}
Hervey~De Saint-Denys.
\newblock {\em Dreams and How to Guide Them}.
\newblock Duckworth Pub, 1982.

\bibitem{saumell2012epidemic}
Anna Saumell-Mendiola, M~Angeles Serrano, and Marian Boguna.
\newblock Epidemic spreading on interconnected networks.
\newblock {\em Physical Review E}, 86(2):026106, 2012.

\bibitem{schweickert2007properties}
Richard Schweickert.
\newblock Properties of the organization of memory for people: Evidence from
  dream reports.
\newblock {\em Psychonomic Bulletin \& Review}, 14(2):270--276, 2007.

\bibitem{serrano2009extracting}
M~{\'A}ngeles Serrano, Mari{\'a}n Bogu{\~n}{\'a}, and Alessandro Vespignani.
\newblock Extracting the multiscale backbone of complex weighted networks.
\newblock {\em Proceedings of the National Academy of Sciences},
  106(16):6483--6488, 2009.

\bibitem{Sun2013srep}
Xiaoling Sun, Jasleen Kaur, Stasa Milojevic, Alessandro Flammini, and Filippo
  Menczer.
\newblock Social dynamics of science.
\newblock {\em Scientific Reports}, 3(1069), 2013.

\bibitem{szell2010multirelational}
Michael Szell, Renaud Lambiotte, and Stefan Thurner.
\newblock Multirelational organization of large-scale social networks in an
  online world.
\newblock {\em Proceedings of the National Academy of Sciences},
  107(31):13636--13641, 2010.

\bibitem{taghva2005arabic}
Kazem Taghva, Rania Elkhoury, and Jeffrey Coombs.
\newblock Arabic stemming without a root dictionary.
\newblock In {\em International Conference on Information Technology: Coding
  and Computing}, volume~1, pages 152--157. IEEE, 2005.

\bibitem{watts1998collective}
Duncan~J Watts and Steven~H Strogatz.
\newblock Collective dynamics of `small-world' networks.
\newblock {\em Nature}, 393(6684):440--442, 1998.

\bibitem{yildirim2007drug}
Muhammed~A Y{\i}ld{\i}r{\i}m, Kwang-Il Goh, Michael~E Cusick,
  Albert-L{\'a}szl{\'o} Barab{\'a}si, and Marc Vidal.
\newblock Drug target network.
\newblock {\em Nature Biotechnology}, 25(10):1119, 2007.

\end{thebibliography}

\balancecolumns
\end{document}